\documentclass[preprint2]{aastex} %a double-column, single-spaced document:

\slugcomment{Accepted to the Astrophysical Journal Letters} 
\shorttitle{Eu abundances in extremely metal-poor stars}
\shortauthors{Ishimaru et al.}

\begin{document}
%======================================================================

\title{Detection of low Eu abundances in extremely metal-poor stars and
the origin of $r$-process elements\altaffilmark{1}}

\altaffiltext{1}{Based on data
collected at Subaru Telescope, which is operated by the National
Astronomical Observatory of Japan.}

\author{Yuhri Ishimaru\altaffilmark{2}, Shinya Wanajo\altaffilmark{3}, Wako Aoki\altaffilmark{4}, and Sean G. Ryan\altaffilmark{5}} 

\bigskip
\affil{accepted for publication in the Astrophysical Journal Letters}

\altaffiltext{2}{Department of Physics, and Graduate School of
Humanities and Sciences, Ochanomizu University,
2-1-1 Otsuka, Bunkyo, Tokyo 112-8610, Japan;
ishimaru@phys.ocha.ac.jp}

\altaffiltext{3}{Department of Physics, Sophia University,
   7-1 Kioi-cho, Chiyoda-ku, Tokyo, 102-8554, Japan;
   wanajo@sophia.ac.jp}

\altaffiltext{4}{National Astronomical Observatory, 
   Mitaka, Tokyo, 181-8588 Japan; aoki.wako@nao.ac.jp}

\altaffiltext{5}{Department of Physics and Astronomy, 
   The Open University, Walton Hall, Milton Keynes, MK7 6AA, United Kingdom;
   S.G.Ryan@open.ac.uk}

\begin{abstract}
We report abundance analyses of three extremely metal-poor
stars with [Fe/H] $\lesssim -3$, using the {\it Subaru} High Dispersion
Spectrograph (HDS). All are found to have sub-solar values
of [Eu/Fe]. Comparison with our chemical evolution model of the Galactic
halo implies the dominant source of Eu to be the low-mass end of the
supernova mass range. 
Future studies of
stars with low Eu abundances will be important to
determine the $r$-process site.
\end{abstract}

%======================================================================
\section{Introduction}\label{sec:intro}

The origin of the rapid neutron capture ($r$-process) elements is still
uncertain. Although a few scenarios such as neutrino winds
\citep{Woosley94} in core-collapse supernovae (SNe), the collapse of
O-Ne-Mg cores resulting from $8 - 10 M_\odot$ stars \citep{Wanajo03},
and neutron star mergers \citep{Freiburghaus99} show some promise, no
consensus has been achieved.

Analyses have shown a large scatter in the abundances of
neutron-capture elements such as Sr, Ba, and Eu among metal-poor stars
in the Galactic halo \citep[e.g., ][]{McWilliam95, Ryan96}.  If the
scatter accurately portrays the state of the interstellar medium (ISM)
from which they formed, large dispersions in excess of
observational errors indicate that the ISM was not fully mixed at the
early epoch of Galactic history and that metal-poor stars contain
products from only one or a few SNe \citep{Ryan91}. The scatter possibly
reflects variation in the yields of SNe from different
mass progenitors; 
the huge dispersions in neutron-capture
elements imply that $r$-process yields are highly
dependent on the masses of SN progenitors.

Eu traces the $r$-process. 
The chemical evolution model of \citet[][hereafter IW99]{Ishimaru99}
explains the large dispersion
of Eu/Fe observed in very metal-poor stars if
the $r$-process elements originate from a limited mass range of
progenitor stars, such as the SNe from stars of $8-10M_\odot$ or
alternatively $>30M_\odot$ \citep[see
also][]{Travaglio99}. 
\citet{Tsujimoto00} have concluded by a similar
approach that the SNe from $\approx 20 M_\odot$ are the dominant source
of the $r$-process elements. A clear difference among these cases is
expected in the stellar distribution of [Eu/Fe] \footnote{
$[X_i/X_j] \equiv \log (N_i/N_j) - \log (N_i/N_j)_\odot$, where $N_i$
indicates abundance of $i$-th element $X_i$.}  at [Fe/H]$\lesssim -3$
(IW99). However, a shortage of observational data in this metallicity
range has made it difficult to distinguish between the proposed 
$r$-process sites.

In this {\em Letter}, we report on three extremely
metal-poor stars which we show to have very low Eu abundances (\S
2). These data are compared with our chemical evolution models to
distinguish between the proposed $r$-process sites (\S 3).

%======================================================================
\section{Observations and Analysis}\label{sec:obs}

We selected three very metal-poor ([Fe/H] $\lesssim -3$) giants,
HD~4306, CS~22878--101, and CS~22950--046, which were known from
previous studies \citep{McWilliam95, McWilliam98} to have [Ba/Fe] $\sim -
1$, typical for their metallicities.  Observations were made with the
High Dispersion Spectrograph (HDS; Noguchi et al. 2002) of the 8.2m
Subaru Telescope in 2001 July, at a resolving power $R=50,000$.  
The total exposure time was 50, 240, and 360 minutes for HD~4306,
CS~22878--101, and CS~22950--046, respectively, resulting in
S/N ratios per 0.012 {\AA} pixel at 4100 {\AA} of
260, 110, and 80.
Data reduction was
performed in the standard way with IRAF\footnote{IRAF is
distributed by the National Optical Astronomy Observatories, which is
operated by the Association of Universities for Research in Astronomy,
Inc. under cooperative agreement with the National Science 
Foundation.}. 
% The S/N ratios per 0.012 {\AA} pixel at 4100 {\AA} in the
% final spectra are 260, 110, and 80 for HD~4306, CS~22878--101, and
% CS~22950--046, respectively.

Equivalent widths were measured by fitting Gaussian profiles to the
absorption lines of species listed in Table~\ref{tab:res}\footnote{The
line list and measurements are available at
http://optik2.mtk.nao.ac.jp/\~{}waoki/EW/}, except for Eu. An
LTE analysis using model atmospheres of \citet{kurucz93}, based on
equivalent widths measured above, was performed for species other than
Eu. Results are given in Table~\ref{tab:res}. For Ba
we included the hyperfine splitting and isotope shifts
\citep{McWilliam98}, assuming isotope ratios of the $r$-process
component in solar system material \citep{Arlandini99}.

The atmospheric parameters (effective temperature: $T_{\rm eff}$~(K),
gravity: $g$~(cm~s$^{-2}$), micro-turbulent velocity: $v_{\rm
mic}$~(km~s$^{-1}$), and metallicity that is assumed to be [Fe/H]) of
CS~22878--101 and CS~22950--046 were adopted from \citet{cohen02} and
\citet{carretta02}, i.e., $T_{\rm eff}$/$\log g$/[Fe/H]/$v_{\rm mic}$ =
4775/1.3/$-3.1$/2.0, and 4730/1.3/$-3.3$/2.0, respectively. We found
that these gravities satisfy the ionization balance between \ion{Fe}{1}
and \ion{Fe}{2}, and the microturbulence velocities lead
to no dependence of the derived abundance on equivalent widths of
\ion{Fe}{1} lines. The derived Fe abundances also agree very well with
the assumed metallicities. So, we adopted these atmospheric
parameters without modification. For HD~4306, we adopted $T_{\rm eff} =
5000$~K based on $V-K=2.15$ \citep{Norris88, Skrutskie97}
%(Norris, Bessell, & Pickles 1988 and 2MASS data: Skrutskie et al. 1997) 
and the temperature scale of \citet{Houdashelt00}.  We derived $g$ and
$v_{\rm mic}$ from the analysis by requiring ionization
balance and no dependence of abundance on line strength.
The resulting parameters, $T_{\rm eff}$/$\log
g$/[Fe/H]/$v_{\rm mic}$ = 5000/2.0/$-3.0$/1.85, are similar to
those of other works \citep[][Honda et al., in
preparation]{McWilliam95}.

\begin{figure}
\includegraphics[width=0.45\textwidth]{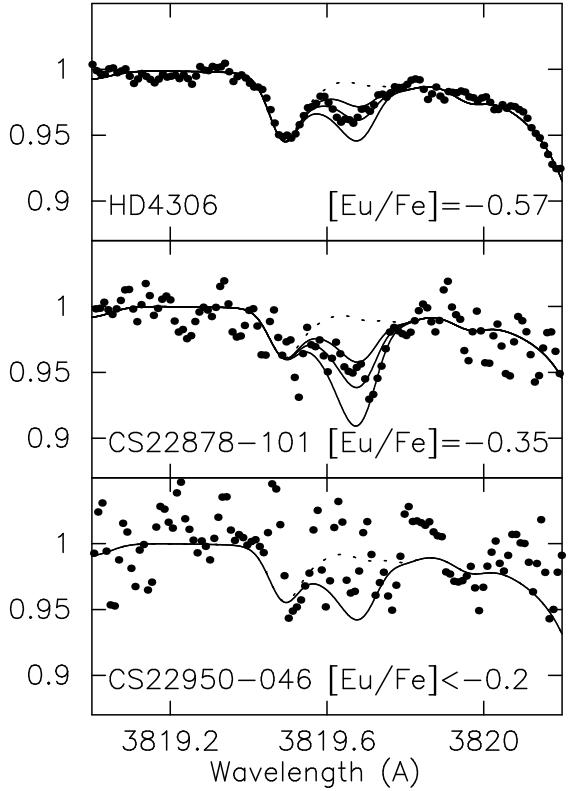}
%\epsscale{0.5}
%\plotone{f1.eps}
\caption[]{Comparison of the observed spectra (dots) and synthetic
ones (lines) near the Eu II 3819.7~{\AA} line. For HD~4306 and
CS~22878--101, three synthetic spectra differ in step of
$\Delta$[Eu/Fe] = 0.2~dex are shown by solid lines. Central line is
calculated for the Eu abundance presented in each panel. The solid
line in the panel for CS~22950--046 indicates the synthetic spectrum
for the upper limit of Eu abundance presented in the panel (see text). 
Dotted lines are the spectra calculated assuming no Eu contribution.}
\label{fig:eu3819}
\end{figure}

For the three \ion{Eu}{2} lines, at 3819, 4129, and 4205~{\AA}, a
spectrum synthesis technique was applied.
The line list includes hyperfine splitting and isotope shifts from
\citet{lawler01}. Contamination by other species was included using the
comprehensive line list of \citet{kurucz95}. Figure~\ref{fig:eu3819}
compares the observed and synthetic spectra for the
3819~{\AA} line, which is the strongest of the three. Eu is clearly
detected in the spectra of HD~4306 and CS~22878--101 but not in
CS~22950--046.  We derived
consistent results from the 4129~{\AA} line, but the
measurement of 4205~{\AA} is uncertain because of severe contamination
by \ion{V}{2} at 4205.09~{\AA}. We adopted the average of the
Eu abundances derived from the 3819 and 4129~{\AA} lines. In
Table~\ref{tab:ew}, we give the equivalent widths of the \ion{Eu}{2}
lines estimated from the spectrum synthesis using the Eu abundances derived
from individual lines. For CS~22950--046, we estimated the upper limit
of its equivalent widths from the 3 $\sigma$ depth of the random noise
level and typical line widths for this object, and determined the upper
limit of the Eu abundance, given in Table~\ref{tab:res}. The continuum
fit is constrained over a wider spectral range that is shown in
Figure~1; a systematic error of 1\%\ would change the [Eu/H] limit from
the 3819~\AA\ line by 0.1.

The random error for each species is estimated from the standard
deviation of the abundances derived from individual lines. The values
are sometimes unrealistically small when the number of lines is small,
so we adopted the larger of the value for the species and that for
\ion{Fe}{1}, 0.08--0.15~dex depending on the species and the quality of
the spectra. We estimated the errors in the Eu abundances by eye from
fitting synthetic spectra to the observed ones, adopting 0.1~dex for
HD~4306 and 0.2~dex for CS~22878--101. The errors in the abundance
determinations due to uncertainties of the atmospheric parameters were
evaluated for $\Delta T_{\rm eff}=100$~K, $\Delta \log g=0.3$, $\Delta
v_{\rm mic}=0.3~$km s$^{-1}$, and $\Delta$ [M/H] = 0.3 for HD~4306, and
added in quadrature with the random error estimated above, to give the
total uncertainty in Table \ref{tab:res}. The larger errors in [Sr/Fe]
reflect the sensitivity to microturbulence 
of the two resonance lines. The [Eu/Fe] values are affected
most significantly by the uncertainty in the surface gravity, by 0.12
for $\Delta \log g = 0.3$.

Comparisons between our results for the two CS objects and those by
\citet{carretta02} show good agreement (within 0.17~dex) for the species
in Table~\ref{tab:res} other than Eu. An exception is Ba 
in CS~22878--101: our abundance is 0.35~dex higher than that by
\citet{carretta02}. The reason for this discrepancy is unclear.  The
equivalent widths of the \ion{Ba}{2} 4554~{\AA} line agree well between
the two studies, and the atmospheric parameters are the same. The
difference in the assumed isotope ratios of Ba in the abundance analyses
do not explain this discrepancy.

%======================================================================
\section{Comparison with the Galactic chemical evolution models}\label{sec:gce}

Assuming that star formation is induced by a single supernova explosion,
IW99 have constructed an inhomogeneous chemical evolution model of the
Galactic halo. The chemical composition of a
newly formed star is a mixture of the products from each SN
and the ISM swept up by the expanding ejecta. This model, along with
some improvements \citep[see][]{Ishimaru03}, leads to a large scatter of
relative abundances for elements that have strong dependencies on the
progenitor masses. The scatters of relative abundances in $\alpha$ and
iron-peak elements, on the other hand, are predicted to be small (as is
observed) owing to the weak dependencies of their yield ratios on the
progenitor masses. We use this model to derive the
Galactic evolution of the $r$-process tracer Eu.
We consider three cases, where the $r$-process elements originate from
SNe of (a) $8-10M_\odot$, (b) $20-25 M_\odot$, and (c) $>30M_\odot$
stars.

Yields for Type II and Type Ia SNe are taken from Nomoto et
al. (1997a, b). The $8-10 M_\odot$ stars are assumed to produce no iron,
since their contribution to the enrichment of iron-peak elements in the
Galaxy is negligible \citep{Wanajo03}. The mass of Eu produced per SN
event is assumed to be constant over the mass range of each case; this
simplification will not affect the result significantly because of the
narrow mass range of Eu-generating SNe assumed in each case. The
requirement that the model reproduces the solar values [Eu/Fe] = [Fe/H]
= 0 implies ejected Eu masses $3.1\times 10^{-7}$, $1.1\times 10^{-6}$,
and $7.8\times 10^{-7} M_\odot$ for cases~(a)-(c), respectively. These
are reasonable amounts from a nucleosynthetic point of view
\citep{Wanajo01, Wanajo03}.

Figure 2 compares the results of these cases with the observational data
from \S 2 (large double circles) and from sources cited in the caption
(small circles). The predicted log number density of stars per unit area
on the [Fe/H]--[Eu/Fe] plane is also shown as a color image. 
%This is
%normalized to fit the metallicity distribution of halo stars
%\citep{Sandage87}.
The thick solid lines indicate the average values of
[Eu/Fe] as a function of metallicity, calculated from the models.  The
solid and thin solid lines indicate, respectively, the 50\% and 90\%
confidence lines.

\begin{figure}
%\epsscale{0.4}
%\plotone{f2.eps}
\includegraphics[width=0.45\textwidth]{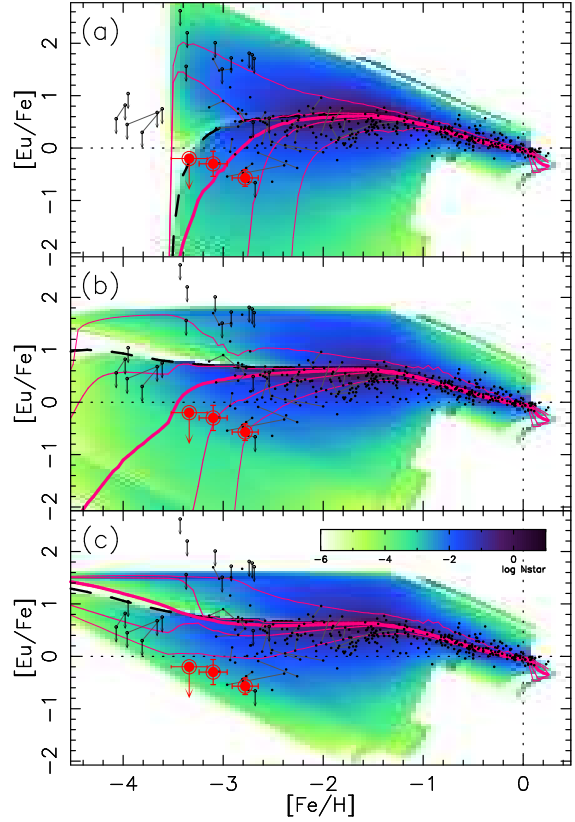}
\caption[]{
Comparison of the observed data with the model predictions.
The $r$-process site is assumed to be SNe of
(a) $8-10 M_\odot$, (b) $20-25 M_\odot$, and (c) $>30M_\odot$ stars. 
The predicted number density of stars per
unit area is color-coded.
The average stellar abundance
distributions are indecated by thick-solid lines with
the 50\% and 90\% confidence intervals 
(solid and thin-solid lines, respectively).
The average abundances of the ISM 
are denoted by the thick-dashed lines.
The current observational data are given by
large double circles, with others
(small circles) taken from
\citet
%{Butcher75, Luck85, Gilroy88, Magain89, 
{
Gratton94, McWilliam95, McWilliam98,
Woolf95, Ryan96, Shetrone96, Sneden96, Westin98,
Burris00, Fulbright00, Norris01, Johnson02a, Johnson02b,
Francois03}; Honda (private communication).
%An $s$-process rich star CS~22898--027,
%an extremely $\alpha$-weak star BD~+80$^{\rm o}$245,
%and carbon rich stars with [C/Fe]$>1$ are excluded. 
}
\label{fig:EuFe}
\end{figure}

\placefigure{fig:EuFe} 
% Comparison of Model and Observation on [Eu/Fe] vs. [Fe/H] 

Figure \ref{fig:EuFe} shows that observable differences between the
cases appear at [Fe/H]$\lesssim -3$.  In case (c), most of the
stars are expected to have [Eu/Fe]$>0$, owing to Eu production solely by
massive, short-lived stars. In cases (a) and (b), significant numbers of
stars having [Eu/Fe] $<$ 0 are predicted at [Fe/H] $\lesssim -3$ owing to
the delayed production of Eu by lower mass SN progenitors. 
This is outstanding in case (a), which has the largest delay.

Most previous observational data, many of them upper limits, distribute
between the 90\% confidence lines for all cases, which has made it
difficult to determine the mass range of the $r$-process site.  Our
newly obtained data add the lowest detections of Eu, at [Fe/H]
$\lesssim - 3$, and help distinguish between the three
cases.  The best agreement can be seen in case (a), in which the three
stars, and most other stars from previous observations are located
between the 50\% confidence lines at [Fe/H] $\lesssim - 3$. In case (b),
all three stars are located below the average (thick solid) line,
although they remain above the 90\% confidence line. In case (c), these
stars are located outside the 90\% confidence region.  We
suggest, therefore, that case (a) is most likely to be the $r$-process
site, i.e. SNe from low-mass progenitors such as $8-10M_\odot$ stars.
More detections of Eu at low metallicity will be
needed to clearly distinguish between cases (a) and (b). 
Note that this result
is not due to a selection bias, since the [Ba/Fe] values
($\sim -1$) are typical of extremely metal-poor stars with [Fe/H] $\sim -
3$ (Ryan et al. 1996).

Our analysis gives [Ba/Eu] values consistent with the solar $r$-process
\citep{Arlandini99} when estimated errors are included (see 
Table~3). Hence our result may hold for heavy $r$-process elements
with $Z \ge 56$, not just $Z \simeq 63$. The values of [Sr/Ba], however,
are significantly higher than for the solar $r$-process,
implying that these three stars exhibit 
light $r$-process elements ($Z < 56$) produced in more massive SNe ($>
10 M_\odot$).

The discussion above suggests that the production of the $r$-process
elements is associated with a small fraction of SNe near the low-mass
end of the range. Neutrino winds in the explosions of massive stars may
face difficulties in being a dominant source of the $r$-process
elements. \citet{Wanajo01} have demonstrated that an $r$-process in the
neutrino winds proceeds from only very massive proto-neutron stars,
which might result from massive progenitors such as $\gtrsim 20-30
M_\odot$ stars, which is similar to case~(b). Hypernovae \citep[$> 20-25
M_\odot$,][]{Maeda03} or pair-instability supernovae \citep[$140-260
M_\odot$,][]{Heger02}, resulting from stars near the high mass-end of
the SN progenitors, similar to case~(c), are clearly excluded as the
major $r$-process site.

We suggest, therefore, that the dominant $r$-process site is SN
explosions of collapsing O-Ne-Mg cores from $8-10 M_\odot$ stars
\citep{Wheeler98}. Recently, \citet{Wanajo03} have demonstrated that the
prompt explosion of the collapsing O-Ne-Mg core from a $9 M_\odot$ star
reproduces the solar $r$-process pattern for nuclei with $A > 130$, and
is characterized by a lack of $\alpha$-elements and only a small amount
of iron-peak elements. This clearly differs from more massive SNe with
iron cores ($> 10 M_\odot$) that eject both these elements, and is
consistent with the fact that the abundances of heavy $r$-process
elements in stars with [Fe/H] $\sim - 3$ are not related with those of
iron-peak elements or of elements with lower atomic numbers
\citep{Qian03}.

This study shows the importance of detecting Eu in extremely metal-poor
stars to explore the origin of $r$-process elements. Further observations
% of extremely metal-poor stars 
are needed to
confirm the origin.
% of $r$-process elements.

%======================================================================
\acknowledgements

This work was supported in part by a Grant-in-Aid for Scientific
Research (13740129) from the Ministry of Education, Culture, Sports,
Science, and Technology of Japan, and by PPARC (PPA/O/S/1998/00658).

%======================================================================
%\placetable{}
%\placefigure{}
%======================================================================

%\clearpage

%======================================================================
%\clearpage

\begin{deluxetable}{lcccccccccccccc}
%\tablecaption{[FE/H] AND RELATIVE ABUNDANCE, [X/FE] \label{tab:res}}
\tablecaption{ELEMENTAL ABUNDANCES \label{tab:res}}
\startdata
\tableline
\tableline
 &\multicolumn{4}{c}{HD~4306} &&\multicolumn{4}{c}{CS~22878-101} && \multicolumn{4}{c}{CS~22950-046} \\
  \cline{2-5}  \cline{7-10}    \cline{12-15} 
     & [X/Fe]$^{\rm a}$ & n & $\sigma$$^{\rm b}$ & $\log\epsilon$ && [X/Fe]$^{\rm a}$ & n & $\sigma$$^{\rm b}$ & $\log\epsilon$ && [X/Fe]$^{\rm a}$ & n & $\sigma$$^{\rm b}$ & $\log\epsilon$ \\
\tableline
%\ion{Mg}{1} & $+$0.64 &  5 & 0.09 & $+$5.44 && $+$0.49 &  5 & 0.11 & $+$4.97 && $+$0.30 &  5 & 0.16 & $+$4.54 \\
%\ion{Ca}{1} & $+$0.50 & 10 & 0.09 & $+$4.07 && $+$0.26 & 11 & 0.10 & $+$3.51 && $+$0.13 &  9 & 0.15 & $+$3.14 \\
%\ion{Ti}{1} & $+$0.42 & 27 & 0.09 & $+$2.58 && $+$0.26 & 26 & 0.11 & $+$2.10 && $+$0.13 & 15 & 0.16 & $+$1.73 \\
%\ion{Ti}{2} & $+$0.48 & 36 & 0.16 & $+$2.64 && $+$0.36 & 37 & 0.18 & $+$2.20 && $+$0.16 & 42 & 0.20 & $+$1.76 \\
\ion{Fe}{1} & $-$2.76 &168 & 0.13 & $+$4.74 && $-$3.14 &161 & 0.14 & $+$4.36 && $-$3.34 &141 & 0.18 & $+$4.16 \\
\ion{Fe}{2} & $-$2.79 & 19 & 0.13 & $+$4.71 && $-$3.07 & 20 & 0.14 & $+$4.43 && $-$3.34 & 17 & 0.18 & $+$4.16 \\
\ion{Sr}{2} & $+$0.28 &  2 & 0.27 & $+$0.42 && $-$0.15 &  2 & 0.28 & $-$0.33 && $-$0.18 &  2 & 0.30 & $-$0.60 \\
\ion{Ba}{2} & $-$1.09 &  2 & 0.14 & $-$1.65 && $-$0.73 &  2 & 0.16 & $-$1.61 && $-$1.31 &  2 & 0.19 & $-$2.43 \\
\ion{Eu}{2} & $-$0.57 &  2 & 0.16 & $-$2.80 && $-$0.30 &  2 & 0.24 & $-$2.85 && $<-$0.2 &  1 & ...  & $<-$3.0 \\
\tableline
\enddata

$^{\rm a}$ [Fe/H] for \ion{Fe}{1} and \ion{Fe}{2}.

$^{\rm b}$ Uncertainty in [Fe/H] or [X/Fe] values. 

$^{\rm c}$ $\log\epsilon (X) = \log_{10} (N_X/X_{\rm H}) + 12.0$ for
 elements $X$. 

\end{deluxetable}

%\clearpage

%\newpage
%======================================================================
\begin{deluxetable}{cccc}
\tablewidth{0pt}
\tablecaption{EQUIVALENT WIDTH OF EU LINES \label{tab:ew}}
\startdata
\tableline
\tableline
Wavelength  & \multicolumn{3}{c}{Equvalent Width (m{\AA})} \\
    \cline{2-4} 
{\AA}   & HD~4306 & CS~22878--101 & CS~22950--046 \\
\tableline
3819.7  & 3.8 & 7.4 & $<$6.8 \\
4129.7  & 2.2 & 5.3 &  \\
4205.0  & 2.5 & 5.0 & $<$3.4 \\
\tableline
\enddata
\end{deluxetable}

%\clearpage

%\newpage
%======================================================================
\begin{deluxetable}{ccccc}
%\footnotesize
\tablecaption{Abundance rations among Sr, Ba, and Eu \label{rat}}
\tablewidth{0pt}
\tablehead{
\colhead{} &
\colhead{HD 4306} &
\colhead{CS 22878-101} &
\colhead{CS 22950-046} &
%\colhead{\mbox{solar\\ $r$-process\tablenotemark{\dag}}}}
\colhead{solar\tablenotemark{\dag}}}
\startdata
{\rm [Sr/Ba]} & $1.37\pm 0.30$  & $0.58\pm 0.32$  & $1.13\pm 0.36$ & $-0.10$ \\
{\rm [Ba/Eu]} & $-0.52\pm 0.21$ & $-0.43\pm 0.29$ & $>-1.11$       & $-0.69$
\enddata
\tablenotetext{\dag}{solar system $r$-process abundances by \citet{Arlandini99}}
\end{deluxetable}

%\clearpage

%\newpage
%======================================================================

%\clearpage

%======================================================================
%\newpage

%\begin{figure}
%%\figurenum{}
%%\epsscale{0.8}
%%\plotone{.ps}
%\plottwo{EuFe1.ps}{EuFe2.ps}
%\caption{}
%\end{figure}

%%======================================================================
%%\newpage
%\begin{figure}
%%\vspace{-5cm}
%\includegraphics[angle=-90, width=0.5\textwidth]{BaEu.ps}
%\caption[]{Same as Fig.~\ref{fig:EuFe}, but for 
%[Ba/Eu] vs.~[Fe/H] relation. 
%The dotted line indicates the solar-system $r$-process value.}
%\label{fig:BaEu}
%\end{figure}

%%======================================================================
%%\newpage
%\begin{figure}
%%\vspace{-5cm}
%\includegraphics[angle=-90, width=\textwidth]{BaFe.ps}
%\caption[]{Same as Fig.~\ref{fig:EuFe}, but for 
%[Ba/Fe] vs.~[Fe/H] relation. 
%The average (the thick lines) and
%LOESS lines (the thin lines) for observational data
%are taken \citet{Norris01}}
%\label{fig:BaFe}
%\end{figure}

%======================================================================

\begin{thebibliography}{}
%\bibitem[Argast et al.(2000)]{Argast00}
% Argast, D., Samland, M., Gerhard, O. E., Thielemann, F.-K. 
% 2000, \aap, 356, 873
\bibitem[Arlandini et al. (1999)]{Arlandini99} 
 Arlandini, C., K\"{a}ppeler, F., Wisshak, K.,  Gallino, R., Lugaro, M., 
 Busso, M., \& Straniero, O., 1999, \apj, 525, 886
%\bibitem[Audouze \& Silk(1995)]{Audouze95}
% Audouze, J. \& Silk, J. 1995, \apjl, L451, 49
\bibitem[Burris et al.(2000)]{Burris00}
 Burris, D. L., Pilachowski, C. A., Armandroff, T. E., Sneden, C.,
 Cowan, J. J., Roe, H.
 2000, \apj, 544 302
%% `Neutron-Capture Elements in the Early Galaxy: Insights from a Large
%% Sample of Metal-poor Giants'
%%\bibitem[Butcher(1975)]{Butcher75}
%% Butcher, H. R. 1975, \apj, 199, 710
\bibitem[Carretta et al.(2002)]{carretta02} Carretta, E., Gratton,
R., Cohen, J. G., Beers, T. C., Christlieb, N. 2002, \aj 124, 481
%\bibitem[Cayrel et al.(2001)]{Cayrel01} 
% Cayrel, R., Hill, V., Beers, T. C., Barbuy, B., Spite, M., Spite, F.,
% Plez, B., Andersen, J., Bonifacio, P., Francois, P., Molaro, P., 
% Nordstrom, B., \& Primas, F. 2001, \nat, 409, 691
\bibitem[Cohen et al.(2002)]{cohen02} 
 Cohen, J. G.,  Christlieb, N., Beers, T. C.,  
 Gratton, R., Carretta, E. 2002, \aj, 124, 470
%\bibitem[Cowan et al.(1999)]{Cowan99}
% Cowan, J. J., Pfeiffer, B., Kratz, K.-L., Thielemann, F.-K., Sneden,
% C., Burles, S., Tytler, D., \& Beers, T. C. 
% 1999, \apj, 521, 194
%% `R-Process Abundances and Chronometers in Metal-poor Stars'
%\bibitem[Depagne et al.(2000)]{Depagne00}
% Depagne, E., Hill, V., Christlieb, N., \& Primas, F. 
% 2000, \aap, 364, 6 
%% `Abundance analysis of two extremely metal-poor stars from the
%% Hamburg/ESO Survey \'
%\bibitem[Fields, Truran, \& Cowan(2002)]{Fields02}
% Fields, B. D., Truran, J. W., \& Cowan, J. J. 2002, \apj, 575, 845
\bibitem[Freiburghaus et al.(1999)]{Freiburghaus99}
 Freiburghaus, C., Rosswog, S., \& Thielemann, F. -K. 1999, \apjl, 525, L121
\bibitem[Fran\c cois et al.(2003)]{Francois03}
 Fran\c cois, P., et al. 2003, \aap, 403, 1105
% Fran\c cois, P., Depagne, E., Hill, V., Spite, M., Spite, F., Plez, B., 
% Beers, T.C., Baubuy, B., Cayrel, R., Andersen,J., Bonifacio, P., 
% Molaro, P., Nordstr\"om, & Primas, F., 2003, A&A 403, 1105
\bibitem[Fulbright(2000)]{Fulbright00}
 Fulbright, J. P. 
 2000, \aj, 120, 1841
%% `Abundances and Kinematics of Field Halo and Disk
%% Stars. I. Observational Data and Abundance Analysis'  
%\bibitem[Gilroy et al.(1988)]{Gilroy88}
% Gilroy, K. K., Sneden, C., Pilachowski, C. A., \& Cowan, J. J. 1988, 
% \apj, 327, 298
\bibitem[Gratton \& Sneden(1994)]{Gratton94}
 Gratton, R. G., Sneden, C. 1994, \aap, 287, 927
\bibitem[Heger \& Woosley(2002)]{Heger02}
 Heger, A. \& Woosley, S, E. 2002, \apj, 567, 532
%\bibitem[Hill et al.(2002)]{Hill02}
% Hill, V., et al. 
% 2002, \aap, 387, 560
% Hill, V., Plez, B., Cayrel, R., Beers, T.C., Nordstr\"om, B., 
% Andersen, J., Spite, M., Spite, F., Barbuy, B., Bonifacio, P., 
% Depagne, E., Fran\c{c}ois, P., Molaro, P., \& Primas, F.
% 2002, \aap, 387, 560
%\bibitem[Hillebrandt, Nomoto, \& Wolff(1984)]{Hillebrandt84}
% Hillebrandt, W., Nomoto, K., \& Wolff, R. G. 1984, \aap, 133, 175
\bibitem[Houdashelt, Bell \& Sweigart(2000)]{Houdashelt00}
 Houdashelt, M.L., Bell, R.A., \& Sweigart, A.V. 2000, \apj, 119, 1448 
\bibitem[Ishimaru \& Wanajo(1999)]{Ishimaru99}
 Ishimaru, Y., Wanajo, S.
 1999, \apjl, L511, 33 (IW99)
%\bibitem[Ishimaru \& Wanajo(2000)]{Ishimaru00}
% Ishimaru, Y. \& Wanajo, S. 2000,  in First Stars, ed. A. Weiss,
% T. Abel,\& V. Hill (Berlin: Springer), 189
% ------------------. 2000,  in First Stars, ed. A. Weiss,
% T. Abel,\& V. Hill (Berlin: Springer), 189
%% `Enrichment of the R-Process Element Europium in the Galactic Halo'
\bibitem[Ishimaru, Prantzos, \& Wanajo(2003)]{Ishimaru03}
 Ishimaru, Y., Prantzos, N., \& Wanajo, S. 
 2003, \nphysa, 718, 671
\bibitem[Johnson (2002)]{Johnson02a}
 Johnson, J. A. 2002, \apjs, 139, 219 
\bibitem[Johnson \& Bolte(2002)]{Johnson02b}
 Johnson, J. A. \& Bolte, M. 2002, \apj, 579, 616
\bibitem[Kurucz(1993)]{kurucz93} Kurucz, R. L., 1993, CD-ROM 13, 
 ATLAS9 Stellar Atmospheres Programs and 2km/s Grid 
 (Cambridge: Smithsonian Astrophys. Obs.)
\bibitem[Kurucz(1995)]{kurucz95} Kurucz, R. L., 1995, CD-ROM 23,  
 (Cambridge: Smithsonian Astrophys. Obs.)
\bibitem[Lawler et al.(2001)]{lawler01} 
 Lawler, J. E., Wickliffe, M. E., Den Hartog, E. A. \& Sneden, C. 
 2001, \apj, 563, 1075 
%%\bibitem[Luck \& Bond(1985)]{Luck85}
%% Luck, R. E. \& Bond, H. E. 1985, \apj, 292 559
\bibitem[Maeda \& Nomoto(2003)]{Maeda03}
 Maeda, K. \& Nomoto, K. 2003, \apj, in press
%%\bibitem[Magain(1989)]{Magain89}
%% Magain, P. 1989, \aap, 209, 211
%\bibitem[Mathews \& Cowan(1990)]{Mathews90}
% Mathews, G. J., Cowan, J. J. 1990, \nat, 345, 491
%% `New insights into the astrophysical r-process'
\bibitem[McWilliam et al.(1995)]{McWilliam95}
 McWilliam, A., Preston, G. W., Sneden, C., \& Searle, L. 1995, \aj, 109, 2757
%%\bibitem[McWilliam 1997]{McWilliam97}
%% McWilliam, A. 1997, \araa, 35, 503
\bibitem[McWilliam(1998)]{McWilliam98}
 McWilliam, A. 1998, \aj, 115, 1640
%\bibitem[Meyer \& Brown(1997)]{Meyer97}
% Meyer, B. S. \& Brown, J. 1997, \apjs, 112, 199
\bibitem[Noguchi et al. (2002)]{noguchi02} 
Noguchi, K., et al. 2002, \pasj, 54, 855
%Noguchi, K., Aoki, W., Kawanomoto, S., et al. 2002, \pasj, 54, 855
%\bibitem[Nomoto(1984)]{Nomoto84}
% Nomoto, K. 1984, \apj, 277, 791
%\bibitem[Nomoto(1987)]{Nomoto87}
% Nomoto, K. 1987, \apj, 322, 206
\bibitem[Nomoto et al.(1997a)]{Nomoto97a}
 Nomoto, K., Hashimoto, M., Tsujimoto, T., Thielemann, F. -K., 
 Kishimoto, N., Kubo, Y., \& Nakasato, N. 1997a, \nphysa, 616, 79
\bibitem[Nomoto et al.(1997b)]{Nomoto97b}
 Nomoto, K., et al. 1997b, \nphysa, 621, 467
\bibitem[Norris, Bessell, \& Pickles(1988)]{Norris88} 
 Norris, J.E., Bessell, M.S., \& Pickles, A.J. 1988, \apjs, 58, 463 
\bibitem[Norris, Ryan, \& Beers(2001)]{Norris01} 
 Norris, J. E., Ryan, S. G., \& Beers, T. C. 2001, \apj, 561, 1034
%\bibitem[Otsuki et al.(2000)]{Otsuki00}
% Otsuki, K., Tagoshi, H., Kajino, T., \& Wanajo, S. 2000, \apj, 533, 424
%\bibitem[Qian \& Woosley(1996)]{Qian96}
% Qian, Y. -Z. \& Woosley, S. E. 1996, \apj, 471, 331
%\bibitem[Qian \& Wasserburg(2002)]{Qian02}
% Qian, Y. -Z. \& Wasserburg, G. J. 2002, \apj, 567, 515
\bibitem[Qian \& Wasserburg(2003)]{Qian03}
 Qian, Y. -Z. \& Wasserburg, G. J. 2003, \apj, 588, 1099
% ------------------. 2003, \apj, 588, 1099
\bibitem[Ryan, Norris, \& Bessell(1991)]{Ryan91}
 Ryan, S. G., Norris, J. E., \& Bessell, M. S. 1991, \aj, 102, 303
\bibitem[Ryan, Norris, \& Beers(1996)]{Ryan96}
 Ryan, S. G., Norris, J. E., \& Beers, T. C. 1996, \apj, 471, 254
%\bibitem[Sandage \& Fouts(1987)]{Sandage87}
% Sandage, A. \& Fouts, G. 1987, \aj, 97, 74
\bibitem[Shetrone(1996)]{Shetrone96}
 Shetrone, M. D. 1996, \aj, 112, 1517
%\bibitem[Shigeyama \& Tsujimoto(1998)]{Shigeyama98}
% Shigeyama, T. \& Tsujimoto, T.
% 1998, \apjl, 507, 135L
%% `Fossil Imprints of the First-Generation Supernova Ejecta in Extremely
%% Metal-deficient Stars'
\bibitem[and 2MASS data: Skrutskie et al.(1997)]{Skrutskie97}
 Skrutskie, M.F., et al. 1997, 
 in The Impact of Large Scale Near-IR Sky Surveys, 
 ed. F. Garzon et al. (Dordrecht: Kluwer), p. 187
\bibitem[Sneden et al.(1996)]{Sneden96}
 Sneden, C., McWilliam, A., Preston, G. W., Cowan, J. J., Burris, D. L., 
 \& Armosky, B. J. 1996, \apj, 467, 819
%\bibitem[Sneden et al.(1998)]{Sneden98}
% Sneden, C., Cowan, J. J., Debra, L. B., \& Truran, J. W. 1998, \apj, 496, 235
%\bibitem[Sneden et al.(2000)]{Sneden00}
% Sneden, C., Cowan, J. J., Ivans, I. I., Fuller, G. M., Burles, S.,
% Beers, T. C., Lawler, J. E. 2000, \apjl, 533, L139
%\bibitem[Sneden et al.(2003)]{Sneden03}
% Sneden, C., Cowan, J. J., Lawler, J. E., Ivans,  I. I., Burles, S.,
% Beers, T. C., Primas, F., Hill, V., Truran, J. W., Fuller, G. M.,
% Pfeiffer, B., \& Kratz, K. -L.
% 2003, \apj, in press (astro-ph/0303542)
%\bibitem[Spite \& Spite(1978)]{Spite78}
% Spite, M., Spite, F. 1978, \aap, 67, 23 
%%  `Nucleosynthesis in the Galaxy and the chemical composition 
%%  of old halo stars'
%\bibitem[Sumiyoshi et al.(2001)]{Sumiyoshi01}
% Sumiyoshi, K., Terasawa, M., Mathews, G. J., Kajino, T., Yamada, S., \&
% Suzuki, H. 2001, \apj, 562, 880
%\bibitem[Takahashi, Witti, \& Janka(1994)]{Takahashi94}
% Takahashi, K., Witti, J., \& Janka, H. -Th. 1994, \aap, 286, 857
\bibitem[Travaglio et al. (1999)]{Travaglio99}
 Travaglio, C., Galli, D., Gallino, R., Busso, M., Ferrini, F., Straniero, O. 
 1999, \apj, 521, 691
%% `Galactic Chemical Evolution of Heavy Elements: From Barium to Europium' 
%\bibitem[Travaglio et al. (2001)]{Travaglio01}
% Travaglio, C., Galli, D., Burkert, A.
% 2001, \apj, 547, 217 
%% `Inhomogeneous Chemical Evolution of the Galactic Halo: Abundance of
%% r-Process Elements'
%\bibitem[Truran(1981)]{Truran81}
% Truran, J. W. 1981, \aap, 97, 391
%\bibitem[Tsujimoto, Shigeyama, \& Yoshii(1999)]{Tsujimoto99}
% Tsujimoto, T., Shigeyama, T., \& Yoshii, Y.
% 1999, \apjl, L519, 63
%% `Chemical Evolution of the Galactic Halo through Supernova-induced 
%% Star Formation'
\bibitem[Tsujimoto, Shigeyama, \& Yoshii(2000)]{Tsujimoto00}
 Tsujimoto, T., Shigeyama, T., \& Yoshii, Y.
 2000, \apjl, L531, 33
%% `Probing the Site for R-Process Nucleosynthesis with Abundances of
%% Barium and Magnesium in Extremely Metal-poor Stars'
\bibitem[Wanajo et al.(2001)]{Wanajo01}
 Wanajo, S., Kajino, T., Mathews, G. J., \& Otsuki, K. 2001, \apj, 554, 578
%\bibitem[Wanajo et al.(2002)]{Wanajo02}
% Wanajo, S., Itoh, N., Ishimaru, I., Nozawa, S. \&  Beers, T. C. 2002,
% \apj, 577, 853
\bibitem[Wanajo et al.(2003)]{Wanajo03}
 Wanajo, S., Tamamura, M., Itoh, N., Nomoto, K., Ishimaru, I., 
 Beers, T. C., \& Nozawa, S. 2003, \apj, 593, 968
\bibitem[Westin et al.(1998)]{Westin98}
 Westin, J., Sneden, C., Gustafsson, B., Edvardsson, B., Cowan, J. J. 
 1998, \baas, 30, 1317
\bibitem[Wheeler, Cowan, \& Hillebrandt(1998)]{Wheeler98}
 Wheeler, J. C., Cowan, J. J., \& Hillebrandt, W. 1998, \apjl, L493, 101
\bibitem[Woolf, Tomkin, \& Lambert(1995)]{Woolf95}
 Woolf, V. M., Tomkin, J., \& Lambert, D. L. 1995, \apj, 453, 660
\bibitem[Woosley et al.(1994)]{Woosley94}
 Woosley, S. E., Wilson, J. R., Mathews, G. J., Hoffman, R. D., \&
 Meyer, B. S. 1994, \apj, 433, 229

\end{thebibliography}
\end{document}